# AI loyalty: A New Paradigm for Aligning Stakeholder Interests

Anthony Aguirre[*], Gaia Dempsey[†], Harry Surden[‡], and Peter B. Reiner[§]

## Abstract

When we consult with a doctor, lawyer, or financial advisor, we generally assume that they are acting in our best interests. But what should we assume when it is an artificial intelligence (AI) system that is acting on our behalf? Early examples of AI assistants like Alexa, Siri, Google, and Cortana already serve as a key interface between consumers and information on the web, and users routinely rely upon AI-driven systems like these to take automated actions or provide information. Superficially, such systems may appear to be acting according to user interests. However, many AI systems are designed with embedded conflicts of interests, acting in ways that subtly benefit their creators (or funders) at the expense of users. Unlike the relationship between an individual and a doctor, lawyer, or financial advisor, there is no requirement that AI systems act in ways that are consistent with the users' best interests. To address this problem, in this paper we introduce the concept of AI loyalty. AI systems are loyal to the degree that they are designed to minimize, and make transparent, conflicts of interest, and to act in ways that prioritize the interests of users. Properly designed, such systems could have considerable functional and competitive – not to mention ethical – advantages relative to those that do not. Loyal AI products hold an obvious appeal for the end-user and could serve to promote the alignment of the long-term interests of AI developers and customers. To this end, we suggest criteria for assessing whether an AI system is sufficiently transparent about conflicts of interest, and acting in a manner that is loyal to the user, and argue that AI loyalty should be deliberately considered during the technological design process alongside other important values in AI ethics such as fairness, accountability privacy, and equity. We discuss a range of mechanisms, from pure market forces to strong regulatory frameworks, that could support incorporation of AI loyalty into a variety of future AI systems.

[*] Professor of Physics, University of California at Santa Cruz and co-founder, Future of Life Institute,

[†] 7th Future

[‡] Professor of Law, University of Colorado at Boulder

[§] Professor of Psychiatry and founder, Neuroethics Collective, University of British Columbia

## Introduction

Conflicts of interest can arise when we try to satisfy our duties to two or more entities. Traditionally, the parties that find themselves in such situations have been individuals or organizations. However, we are witnessing a radical shift in terms of the players that might be involved: in the modern world, artificial intelligence (AI)[**] systems introduce a new set of stakeholder dynamics where conflicts of interests arise. For instance, a user who searches for a product using an AI system might assume that the results that are returned are the most relevant, highest quality, or best value, but in fact, such systems often prioritize results that provide the most financial benefit to the software designers, and algorithmic news feeds may prioritize user engagement time and advertising dollars over users' desire for true and useful information[1]. As the capability and sophistication of AI engines providing recommendations, making decisions, and taking action visibly and invisibly within the complex web of corporate and consumer stakeholders mature, these misalignments in interest can be expected to grow in importance. The issue is especially problematic when we consider the power and knowledge imbalance between transnational technology companies with financial resources greater than that of many sovereign governments[2], and individual users. This article aims to address the issue of conflict of interest in AI software through the lens of the concept of **AI loyalty.** While the framework that we develop may have widespread applications for AI systems more generally, we begin with the implications of AI loyalty in products that offer AI-based services to individual consumers. For this reason, we explore the concept through the particularly salient context of AI personal assistants, in which an AI system acts as an agent that makes recommendations and takes actions on behalf of a user in a way that echoes a human assistant or consultant.

## Contemporary AI personal assistants

Virtual assistants powered by AI are becoming a ubiquitous presence in the modern world. Systems such as Apple's Siri, Amazon's Alexa, Google's Assistant and Microsoft's Cortana can be found on billions of smartphones, smart speakers, and digital buttons and are capable of handling a growing array of digital tasks. From the users' perspective, these assistants generally present themselves as performing tasks or providing information *for the user's convenience and benefit*, without making their limitations or conflicts of interest explicit. However from a business perspective, they may play various roles: as a means of harvesting data that can be used to form digital profiles to inform future product development, target advertising, or feed machine learning systems, as a value-add to a hardware system, as an interface to increase the purchase of particular products, and so on[3].

Over the last 20 years, business practices enabled by accelerating technological advances in consumer tracking have changed long standing assumptions about conflicts of interest between businesses and customers. Traditionally, users purchasing a product or service were safe in assuming that the service would align with the user's interests in exchange for the financial compensation they paid. But this economic arrangement and its attendant assumptions have been undermined by the new class of widely available digital products, whose replication and distribution costs are negligible in comparison to that of physical goods. Today, businesses provide technological services to millions or billions of users at no direct charge, and monetize this relationship through less obvious pathways[4]. These alternative pathways include the gathering and analysis of personal consumer data, aggregation and sale of that data, and use of that data to influence consumer purchasing, capture engagement time in order to sell ads, or other personal actions. Importantly, these monetization strategies are often not apparent to end-users. Worse, companies often take active steps to obfuscate these activities from consumers; and many of these same companies are developing the most widespread AI assistants.

---

[**] We use "AI" here as a catchall term to include a variety of machine-learning and related techniques incorporated into contemporary systems.

This represents a potential divergence of interest that has garnered increasing attention and concern. A virtual assistant may, without disclosure, encourage (or only allow) purchases from a particular vendor or otherwise serve its home company's interests[5]. Moreover, the data-gathering business model has been heavily critiqued as being ethically fraught. Companies' financial incentive is to extract ever increasing amounts of personal information regardless of benefit to the user.[††] As a result, consumer backlash and accusations of manipulation, privacy violation and more have been levelled at technology companies[6,7,8]. In an attempt to manage the situation, governments have imposed historically heavy fines for breaches of privacy regulations both in the US[9] and in the EU where GDPR rules hold sway[10].

We propose to turn this situation on its head, beginning with an ethical view of how our relationship to AI assistants should look, from the perspective of the consumer. We will argue that a key feature of a new, ethically sound product category for AI assistants is that the assistant should be loyal to the consumer – that it should consider the user's interests first and foremost. Importantly, this loyalty principle applies to consumer AI systems more broadly, and not just to AI assistants. AI assistants are a specific, tractable example of a system where the AI loyalty framework can be used to shed light on and ameliorate analogous problems that are endemic in AI systems more generally and will become higher stakes as these systems' capabilities increase. In our recommendations section, we will discuss the range of factors that would need to be present in order to command a sea change in the stakeholder mindsets and ultimately widespread development and adoption of loyal AI assistants, including market demand, regulatory pressure, consumer advocacy, and media attention.

## What might we want AI assistants to do?

The current crop of AI assistants, even if quite limited as compared to the promise of coming years, are reasonably capable. Voice recognition software allows users to communicate with them using natural language, and they can answer factual questions, take dictation, send texts and emails, schedule events, make recommendations and control the expanding suite of "smart" devices taking up residence in the home, car, and office (the market share of which tech giants and myriad startups are vying for).

With time, the capabilities of AI assistants are likely to increase such that they become more and more like *bona fide* executive assistants[11] with some tasks carried out autonomously without user oversight and others – in particular higher stakes scenarios that involve greater complexity – requiring significant user input. Even without invoking the development of full-fledged artificial general intelligence, it seems plausible that one day AI assistants could be involved in making significant purchases[12], booking travel reservations, negotiating the purchase of a home, supporting detailed logistics and event planning, and supporting the decision-making process on topics related to questions of health, nutrition, childcare, energy consumption, the job market, investments – anything that an individual would ask a competent operator or highly intelligent advisor for support on[13].

---

[††] Such systems can act not only to extract data but also to deliberately increase reliance on the ecosystems of products they connect with. Because of this, negative feedback loops can result that decrease the agency of the individual yet lock them into an ecosystem that is all but necessary for their participation in large swathes of the global economy and social fabric; see Maldoff G & Tene O, The Costs of Not Using Data: Balancing Privacy and the Perils of Inaction  15 J.L. Econ. & Pol'y 41 (2019) and Siemoneit A, An offer you can't refuse _ Enhancing personal productivity through 'efficiency consumption', ZOE Discussion Papers (2019).
.

AI assistants could protect users from an array of cyberattacks, warding off not just spam but also phishing, malware, and even legal but exploitative scams. They could also serve as a central repository and negotiator of user preferences on privacy and similar matters, communicating those preferences to a system with which the user is interacting, and/or warning a user when a system does not respect those preferences. Similarly, users' ethical preferences (for example to favor or avoid particular industry practices or company qualities) could be folded into recommendations to the user.

On a more aspirational level, an AI assistant could help users achieve a better state of well-being – a software layer that users can cooperate with to help them more effectively steer, drive, and course-correct along their personal path of self-improvement. Developers are already releasing AI assistants tasked with helping users improve their health and material welfare, gently nudging[14] them towards better nutrition, serving as AI therapists, and helping them save for a rainy day[15]. Even more ambitious would be programming that helps users further develop the skills that provide meaning to the human experience: the satisfaction that derives from accomplishment, the value of social connectedness, the nourishment of one's inner life. AI assistants could "know" when and how it is more appropriate to encourage the user to reflect on the values underlying their habits and teach them to evaluate their lives' competing demands[16]. Designing AI assistants with these features would go some distance in lessening the worry that over-reliance upon AI assistants for instrumental tasks might enfeeble us, diminishing our capability to effectively navigate the world[17]. The concept of AI loyalty becomes of paramount importance in systems with such a high degree of access to the user's sense of identity, well-being, and general mental and emotional state.

## Aligning interests – the case for loyal AI Assistants

For most people, using an AI assistant such as Siri or Alexa simply involves speaking into a device and hearing a response. While the small print in the Terms of Use generally make it clear that collecting and exploiting user information is a condition of gaining access to the service, most users blithely ignore the Terms of Use entirely[18]. The result is often relatively unfettered collection of user information, setting the stage for AI-driven algorithms to persuade, cajole and manipulate user behavior[7,19] - all without much incentive or any requirement to consider the best interests of the user. We suggest that loyal AI Assistants – AI agents that explicitly put the interests of the user at the fore – represent an appropriate resolution of this problem, and a significant marketing advantage, provided they are implemented in a trustworthy manner[‡‡]. Loyalty of this sort represents a version of the value alignment problem – the challenge of ensuring that an AI-driven agent act in accord with some set of values. We will not recapitulate the full set of arguments that have been proffered on this topic but note that its solution is fundamental to the development of the next generation of AI[20,21].

The idea of "loyalty" is well established in human professional relationships in the form of the fiduciary duty owed to their clients by physicians, therapists, and lawyers[22]. We expect these fiduciaries to put our interests first and to carefully and ethically handle any conflicts of interest that may arise. In turn, we feel comfortable sharing important and personal information and trusting their recommendations. AI loyalty would logically be even stronger than that of a human fiduciary, as unlike a doctor or lawyer the AI presumably has no interest (for example salary, pride, etc.) of its own[§§]. The wellbeing of the end-user could be absolutely paramount in terms of recommended products or purchases. In terms of actions, depending upon the business model employed (see below), a loyal AI assistant could be configured such that it was primarily responsive to the interests of the user so long as actions that it might be asked to provide are not illegal. (Even here exceptions could obtain. Imagine a scenario in which one is rushing to the hospital in a self-driving car, and the user asks the AI assistant to go 5 miles per hour above the speed limit. The AI assistant might respond by reminding the user that speeding is against the law, and the user could accept responsibility for the legal infraction.)

---

[‡‡] It might be objected that conflicts-of-interest between AI users and developers should simply be transparently declared, and that this information is sufficient for users to fold into their decision-making. But in a system that has such conflicts built-in with some requirement to declare them, the incentives most naturally lead to the declarations either being hidden in lengthy, abstruse, universally unread terms-of-use agreements# or – where this is disallowed – so ubiquitous as to be simply wasteful and dismissed without effect, as in endless GDPR notices. Neither case leads to meaningful mitigation of the conflict.

A well-functioning loyal AI Assistant – preferably functioning in the context of a clear and robust legal and regulatory framework – would reinforce the possibility of trust between user and algorithmic system. Assuming that the duty of loyalty is maintained over time, one can imagine the relationship moving towards the thick sort of trust that develops in healthy interpersonal relationships, in which the parties have a shared history that reinforces mutual confidence in their actions[15].

This is important not just on a social level but also to improve the overall function of interactions between systems. The deeper the trust that users ascribe to their loyal AI assistants, the more freely they will share information with it. As a result, the loyal AI assistant will have more robust information at its (virtual) fingertips and therefore be better able to discharge its duties towards the user. This virtuous circle is exactly the opposite of the vicious circle we see underway in the current regime, where people who distrust social media are encouraged to be careful about what they share[23], resulting in less robust images of who they are. Trust allows the AI system not just to receive information but also to carefully, appropriately, and where applicable, anonymously *share* information, opening the possibility of new functionalities. Modern anti-spam systems, for example, overpower spam generators by sharing information about what is labeled as spam[24]; loyal AI assistants, similarly, could help their user coordinate with others with common interests. This could include identifying and mitigating low-value drains on users' time and money, as well as protecting the user against manipulative tactics. A loyal AI could also be entrusted to identify (and if so requested, take advantage of) opportunities for its user that are genuinely in the user's interest.

A well-established system of loyal AI assistants would also have significant wider social implications. It is already known that loyalty promotes trust in data stewards[25]. In order to garner the trust of users, loyal AI Assistants would need to satisfy other features including, at a minimum, the ability to explain their actions in easily understandable terms, to maintain data security, and, importantly, be able to communicate with outside entities and agents without compromising privacy. Models for the latter include the well-developed examples of differential privacy[26] and contextual integrity[27], or the federated approach of the privacy-preserving open-source AI assistant framework Almond[28]. The idea of AI loyalty also aligns closely with that of "human compatible" AI[20], in which AI systems aim to accomplish human goals rather than their own (even if human-provided) goals.

---

[§§] We leave aside the issue of whether future advanced AI systems *could* meaningfully have their own interest, but suggest that even if it were possible to create such systems, AI assistants should probably not be among them.

While loyalty in many ways may appear an unalloyed good, there are many subtleties, as in all human interactions. For example, how should the user's interests be balanced against interests other than those of the AI (which presumably do not exist) or its parent company? An objection to the loyalty requirement has been raised by AI pioneer and practitioner Stuart Russell who rightly points out[20] (in the context of quite advanced AI assistants) that an AI might follow the law but still unscrupulously trespass social norms in the process of executing its loyalty obligations to its user. He posits a situation in which the user has accidentally double booked for dinner – both with his wife for their twentieth anniversary and the secretary-general who is flying in for the occasion. As a solution to the dilemma, the AI causes the secretary-general's plane to be delayed. Russell's solution is for the AI to follow a set of moral rules known as consequentialism. We suggest that true loyalty means reflecting the user's values, their social etiquette, and their general preferences about navigating the world. For simple AI assistants such as those available today, such regard would need to be instantiated directly into the system, implicitly consented to by the user when adopting the system. More advanced systems with a greater action space could include, for example, user-definable settings that characterize how to weigh the user's interest against others', with some required minimum weight accorded to the latter. Very sophisticated systems, having a detailed model of their users' preferences, could learn from their user's decisions how to accord weight to the interests and preferences of others, with some hard ethical limits to the loyal AI assistant's behavior in place. That is, an advanced loyal AI assistant would include the consideration of others in its recommendations and actions specifically *because its user would.* However, this arrangement would do nothing to mitigate the amoral actions of bad actors. Such individuals would likely use the power of AI to further their amoral interests – this is an inevitable consequence of putting the growing power of AI at the service of people who enjoy significant liberty in their actions. New legal and social norms will develop to address this reality, often driven by notable/newsworthy transgressions. On the other hand, in the case of people *trying* to do the right thing, AI assistants with such advanced capabilities could even help better align the actions of the user with the user's own moral compass: anytime the user makes a request that appears to transgress what the loyal AI assistant understands to be the morally appropriate thing to do, it could prompt the user to reflect on the issue and await further instructions. Small nudges such as this are already in place in the AI that monitors offensive posts on Instagram[16], and could easily be expanded upon and deployed more widely.

Criteria for near-future Loyal AIs

Although some loyal AI capabilities will require future technical advances, this framework applies to contemporary AI assistants. Here we sketch some principles and criteria for designating an AI assistant system "loyal."

***Value alignment.*** To the extent that the system acts independently of the user to facilitate the completion of some action or task:

- The system should be deliberately designed to eliminate clear conflicts of interest – e.g. AI systems taking automated actions or providing prioritized recommendations that financially benefit the creators or funders of the AI system;
- When clear conflicts of interests do exist, the system should transparently and saliently indicate to users the presence of such conflicts;

- The system's underlying operational criteria and goal (utility) functions should, at minimum, be made transparent so that users (and/or auditors) can determine whether they are in alignment with the user's own goals;
- Preferably, the criteria and goals should be adjustable in terms of balancing tradeoffs in the prioritization of optimizable factors (e.g. price, speed, cost, privacy, etc);
- Optimally, the values utilized by loyal AI should be derived from revealed preferences[21], learned directly from the user where possible[***], appropriately and efficiently requesting user input as needed.

Maintaining value alignment between an AI system and its user is a highly nontrivial problem[29,30] but also an area of active investigation[21,31,32,33]. As discussed below, AI assistants may be a useful proving-ground for alignment techniques.

***Decision transparency.*** To the extent that decisions are made independently of the user, the decision-making process should be transparent and explainable. The system should be designed to empower and include users in decisions, educating the user on the relevant factors forming the basis for those decisions. This accomplishes three important objectives: first it conditions and trains users in the powers and limitations of the system[34], second, it protects against a form of learned helplessness that has been termed digital resignation[35], and third, it facilitates human intervention in the case of automated decision failures.

***Data integrity.*** The system should be aware of the provenance of data and attribute the appropriate legal and privacy rights to its originator. For example, the system should be capable of tracking the origin of data that is being monetized and directing payments (or other value commensurate with that of the data) to the correct entities. Or, in the case of sensitive information such as medical data, the system should be capable of keeping track of which data may legally be shared with which party and use appropriate encryption or other privacy technologies to ensure the right protections are in place[†††].

***Personal Privacy.*** The system should have extreme regard for privacy, including both how and why it retains user data, and how and why it shares user data as appropriate (to accomplish a given task, for example, in an anonymized and encrypted method allowing network effects with other AI assistants.) In addition, the system makes explicit to the user the privacy risks of any particular action, when appropriate.

## Legal framework

To what extent, if any, should the law play in encouraging or requiring AI loyalty? One possible approach involves limited (or no) government involvement. One could imagine a form of industry self-policing that might arise in which the corporate creators of AI systems voluntarily develop and implement AI loyalty best practices during the process of technological design. The possibility of industry self-regulation is not completely far-fetched. There have been past examples of such self-regulatory efforts in the area of privacy and data anti-discrimination. These voluntary, non-governmental regulatory efforts have arisen largely as a reaction to public criticism and as a means to head off pending government regulation. However, once implemented, such voluntarily self-policing efforts are often criticized as half-hearted or self-interested, designed in such a way to placate critics and garner positive publicity, while doing little of substance.

---

[***] This fits the existing machine learning framework of Inverse Reinforcement Learning; see Arora S & Doshi P, A Survey of Inverse Reinforcement Learning: Challenges, Methods and Progress, https://arxiv.org/abs/1806.06877 and, addressing longer term issues, S. Russel, "Human Compatible", Viking, 2019.

[†††] As a salient example, it is very likely that many consumers would happily share substantial personal data with an agency like the U.S.'s Center for Disease Control *and it alone* during a pandemic; this could prove extremely helpful both individually and collectively. But there is currently no trustable mechanism for doing this.

Another possible approach with little or no government role involves regulation through "market discipline." Market discipline is the concept that economic competition (as opposed to government regulation) can bring about some desired social goal when implementing that goal actually confers a competitive advantage. In this vein, one could imagine consumer rights groups desiring or demanding that consumer AI systems become transparent, free of conflict of interest, and loyal to user interests. If such consumer demand were to arise, firms could voluntarily implement and advertise AI loyalty as a market distinguisher for their products. That could result in increased market share and economic advantage over competitors who produce AI products with embedded conflicts of interests, producing competitive pressure that could cause AI loyalty to diffuse more broadly. However, such a "market-discipline" scenario might not actually transpire, as it would require both consumers and producers to value AI loyalty (over other features) such that it would shift their usage preferences in a meaningful way. By contrast, history has shown that consumers often prefer other factors such as speed, ease of use, or low or free price, even at the expense of the concepts such as transparency or privacy that are related to AI loyalty – or consumers may not recognize the importance of these factors until their absence is so baked-in to large-scale systems as to be both problematic and difficult to change. Thus while loyalty may be a significant product differentiator, this market discipline driver is unlikely to suffice on its own to create the conditions necessary for widespread adoption.

There are other approaches in which the law might take a larger role in promoting AI loyalty. Most obviously, Congress or state governments could introduce legislation requiring companies to adopt AI loyalty principles in some way. This could include anything from a requirement that creators of AI systems consider issues of AI loyalty during the product development pipeline, to rules that AI systems disclose and make transparent and obvious clear conflicts of interest. Other regulations might insist upon clear and transparent explanations for how automated results were produced. We see analogs to these requirements in the privacy realm, where recent regulation from California (the California Consumer Privacy Act) and the European Union (The General Data Protection Act) contain similar requirements with respect to privacy and personal data. This approach seems reasonably promising, as there has been much discussion (and proposed legislation) at the state and federal level concerning the regulation of Artificial Intelligence more broadly. The additional step here would be to simply include "AI Loyalty" as an additional principle to be considered in such regulation.

However, some obvious problems that will arise concern the difficult nature of coherently defining, providing standards for, and applying concepts such as "conflict of interest," "best interest," and "loyalty" in the legal context of AI automation. This will raise difficult practical questions such as: how do we actually determine or measure a user's "best interest", which user interests are we measuring, what if one user's interests harm others, and what if an individual user has multiple interests that conflict with one another? Similarly, one could imagine legally distinguishing issues of "AI disloyalty", where AI systems are clearly acting out of conflict of interest (i.e. a system that makes a sub-par automated decision for a user that results in a financial kickback for the AI creator), from more difficult "AI loyalty" issues where user interests are simply under-specified, or are not maximally satisfied by AI systems.

In a different vein, principles of AI loyalty could be viewed through a consumer protection lens. Arguably, AI systems that seem to produce accurate and relevant actions for users but that are actually acting in the interests of others could be characterized as deceptive or unfair trade practices. In this legal framing, there are already numerous federal agencies (such as the Federal Trade Commission), and state agencies (such as State Consumer Protection Offices) that could promulgate regulations, or take enforcement actions against AI systems that engage in controversial, but non-transparent, conflicts of interest against consumers.

Contract law might also play a role in promoting AI loyalty. In some cases, vendors or users of AI assistants may promulgate explicit terms of service or policies governing their use and behavior. In certain circumstances, one could imagine explicit contractual promises that AI assistants will make reasonable efforts to implement systems that promote loyalty, privacy, explainability, and data security. However, unless specifically required by law, such explicit contractual promises do not appear to be a promising avenue due to the difficulty in defining these standards, the subjective nature of these criteria, and the risk of abuse by opportunistic litigants. Relatedly, one could imagine the courts creating common law contractual duties of implied AI loyalty, much in the way that consumer goods have implied warranties of merchantability even in the absence of explicit warranties. Finally, legal rules governing AI loyalty might develop incrementally through consumer class action litigation routes.

A distinct aspect in which the law might play a role concerns liability for the actions of AI systems that have embedded conflicts of interest. Should the creators of AI systems be liable for automated decisions that are clearly suboptimal for users, and if so, to what extent? Perhaps more forward-looking, what happens if AI systems innocently, but autonomously, act on behalf of users in ways that hurt the financial, legal, or physical interests of others? Should the creators of AI systems be liable for such actions? Should the users be liable? What if AI systems are knowingly directed to autonomously act in a way that will likely violate the law? Who should bear the civil or criminal liability for such actions? These critical open questions remain around the intersectionality of loyalty in AI, from both an ethical and a legal perspective.

### Paying for loyal AI assistants

The concept of loyal AI Assistants builds upon, but is distinct from the proposal that large technology companies ought to be regulated as information fiduciaries[36]. Indeed, the information fiduciary model has been critiqued for condoning the tech titans in maintaining their existing business model, collecting user data to drive microtargeting of advertising[37], thereby perpetuating conflicts of interest. Loyal AI demands modification of this business model. At the same time, loyalty could be a major product differentiator and competitive advantage, particularly as privacy and related concerns continue to grow and potentially come under stronger regulatory sway.

It is likely that any model in which the AI system provider has a financial model with revenue resulting from the *way* the loyal AI Assistant *operates* is likely to lead to conflicts between revenue maximization and loyalty to the user. This conflict can be removed if the revenue is tied to the fact that the assistant is used, rather than what it actually does.

One option, therefore, would be for a loyal AI to be bundled with another product, consistent with a business model in which 'turning the flywheel' represents a path to success[38]. For example, Apple's AI assistant Siri is incorporated into all iPhones. Offering a loyal version of Siri would represent an added inducement for people to purchase iPhones, with Apple incorporating the cost of the AI functionality and ongoing support and development in the purchase price of the device and their ongoing support and service plans, such as Apple Care and iCloud. Whether introduced by Apple or its competitors, one can imagine a scenario in which a loyal AI assistant would confer significant competitive advantages to its producers. If successful, such a consumer product would create social and market pressure that could modify the unhealthy balance of power that exists today in the industry.

An alternative model might involve subscription pricing. Basic versions of a loyal AI assistant could be offered for a modest cost, with capability enhancements available as the equivalent of in-app purchases. These are only the most obvious solutions, but the power of entrepreneurial ideas will undoubtedly lead to alternative approaches, each competing for market share in their own way. Some of these business models may come with their own problems – for example there is the worry that if AI assistants with a wide range of capabilities are available at a wide range of cost points, loyal AI assistants might be available most easily only to those with the means to pay for them, widening the gap between the haves and the have-nots.

## Power asymmetries and coordination

Our modern internet-mediated economy embodies an unprecedented asymmetry of information, and in some ways power, between large companies that can collect and process information from many people and enact decisions that affect many people at once, and individuals who must each make decisions and take action based on their own information. Governments and the legal system (including class-action lawsuits) can and do protect citizens from physical harm and (at some level) fraud etc.; but what protection is available to people against pervasive manipulation, privacy invasion, false (and too-numerous) choices, etc.? The traditional model – successful in many ways – has been that informed consumers and citizens "vote with their wallet" and target their resources toward companies and products that serve them best. But the modern information economy is arguably so complex and information-asymmetric that individuals cannot possibly make rational informed decisions reflective of their own interests, values, and objectives without a level of time, effort and expertise that is unavailable to the vast majority of people. This trend is unlikely to change unless a new ingredient is added that helps tip the balance back in favor of autonomous individuals, by providing them with trusted tools that can analyze and navigate a high-complexity economic environment, as well as allow coordinated action to counteract the power of corporations and nation-states. Loyal AI assistants are an example (but only one) of such a new ingredient.

## Relation to the issue of AI alignment

The alignment of AI assistants with their users' interest is a special case of a wider problem. As AI systems become more capable it will make sense to cede more decisions to them. But how can these decisions then be assured to be consonant with the goals and values of their operators, so that it makes sense for an individual (or organization) to delegate these decisions or trust these recommendations? Numerous AI experts among others have pointed out the profound and fundamental difficulty of this problem[20,39,40] arguing that any explicitly specified set of objectives for an agent operating over a very wide action space are virtually guaranteed to lead to unforeseen negative side effects. Just as with human assistants, AI assistants will make mistakes – both in failing to do what they are trying to do, and in trying to do the wrong thing. This will make AI assistants challenging as products, because many users employing different and customized AI assistants for various tasks will lead to instances of users pushing AI assistants in potentially problematic directions, and instances of users objecting to (i.e. complaining about) things the assistants do.  This, however, is an opportunity as well, since through this process the framework of the assistants and their design underpinnings will be repeatedly and adversarially tested and improved, starting when the stakes are low.   This seems much more likely led to robustly aligned AI architectures than for AI systems with limited interaction with a limited number of users and may very well significantly contribute to the broader challenge of AI alignment.

The type of loyalty discussed here is not the only form of AI alignment. Loyalty as such need not be to an individual person.  We may imagine AI systems designed to be loyal to a group, a legal entity such as a corporation or government, a nation, or even to humanity as a whole.  As with human affairs, it may be challenging to determine how to aggregate the preferences of multiple entities within the group, and in those cases the label of loyalty would fit less well, even for systems that are explicitly designed to behave according to agreed upon moral tenets.

## Recommendations

The aim of this paper has been to bring attention and conversation to the importance of loyalty – or lack thereof – in current and future AI systems, focusing on the example of AI assistants.  To the extent that loyalty is desirable, what could make it the norm? Commonly, issues of public concern (e.g. climate change) are only considered by powerful economic players when a combination of market incentives, government regulation, consumer demand, consumer advocacy, media pressure, and shareholder activism are present – no single ingredient alone is sufficient to catalyze meaningful, widespread change in corporate operations and policy. Thus, we offer a diverse set of initial recommendations toward the end of AI loyalty:

- AI researchers and developers should continue to devise, develop, and prototype systems and protocols necessary for loyal AI systems, including private information storage and appropriate sharing, data provenance tracking, decision and goal transparency, and conflict of interest monitoring.
- AI companies in or adjoining the market of personal assistants should consider loyalty as a potential design feature for their existing products and consider developing products with loyalty at their core.
- Major tech companies offering personal assistants should consider loyalty alongside properties like privacy and discrimination. Moreover, they should create internal processes that empower technologists and other product development specialists) to explicitly query the issue of conflict of interest throughout the development pipeline. Companies for whom AI loyalty is a natural part of their product offering should leverage this fact for market advantage.
- Policymakers should consider loyalty alongside other pro-social AI system attributes such as transparency, non-discrimination, privacy, and safety.  At minimum, transparency and disclosure of conflicts of interest should be strongly considered as part of any set of strictures placed on consumer-facing AI products.
- Consumers, while demanding appropriate levels of effectiveness, privacy, safety, and fairness from the products they use, should consider how important AI loyalty is to them, and exhibit this both explicitly and in purchase/use preferences.

## Conclusion

The past 20 years have seen the explosive growth in the power and capability of online platforms that mediate the connection between users and nearly every important part of the social, economic, intellectual and even natural world.  The power and reach of these platforms have created enormous convenience for many users and increasingly are becoming an indispensable part of modern life. However, their advertising-focused ethos and the nearly regulation-free model in which they have been developed "fast" without concern for "breaking things" has resulted in a system with fundamental drawbacks.

Many of these drawbacks are embodied in current-day AI assistants. There are many instances in which AI systems appear to be acting in the users' best interests, but in fact do not do so.   Rather, these systems are subject to either deliberate or accidental technical designs that promote the interests of the system creators (or others) to the detriment of users.   The absence of what we have called "AI loyalty" has largely flown under the radar – as compared for example to privacy and equality – for two reasons. First, although AI systems make automated decisions on behalf of lay users today, these automated decisions tend to be fairly inconsequential data retrieval tasks, such as playing music, creating task reminders, or returning results for search queries about product purchasing or navigation.   Second, the technology underlying contemporaneous AI systems tends to be relatively limited, thereby limiting their scope and reach.

This will change.  When an AI system with an embedded conflict of interest subtly promotes one less attractive product at the expense of another, the impact is relatively small. But if one considers the widespread adoption of AI in the analysis of job interviews, in the  automated assessment of banking and credit applications, in medical diagnostics, and as a (widely derided) means of parole assessment, it becomes clear that AI systems are rapidly being implemented in high-stakes arenas. It is only a matter of time before AI systems develop sufficient technological capabilities that they will be assisting consumers with more consequential decisions, such as assisting parents in finding the best daycare centers for their children, or helping doctors prescribe medication, or managing wealth portfolios independent of human input for long periods of time. These are precisely the scenarios for which we must avoid conflicts of interest. It is therefore important to open up the conversation into AI loyalty today, while the stakes and capabilities remain comparatively low, and to explicitly incorporate considerations of AI loyalty into the technological design and deployment process. In the paper we have outlined criteria to be considered in doing so.

We have also emphasized that beyond mitigating conflicts of interest that are detrimental to users, AI loyalty presents an opportunity: genuinely trustable AI systems can provide services – for example using highly private or proprietary data – that could not (or at least certainly *should* not) be provided by potentially disloyal ones.

In the years to come, the role of AI-based software agents in the affairs of humans will grow very substantially.  We suggest that the current  trajectory in which AI systems grow increasingly capable and versatile while harvesting (and selling) immensely detailed user profiles, providing opaquely sourced and reasoned recommendations, and acting in the interest of the user only insofar as it aligns with the overall growth and profit goals of the system's parent company runs the risk of leading to ever more dystopian outcomes. We can also imagine, however, a trajectory in which individual autonomy is dramatically empowered and overall well-being significantly improved, by loyal AI systems that users can, will, and *should* trust.

We thank Jared Brown, Richard Mallah, Adrian Byram, and Imre Bard for comments on an earlier version of this manuscript.